\documentclass[twocolumn,prl]{revtex4}
\usepackage{graphicx}
\usepackage{subfigure}
\usepackage{bm}
\usepackage{amssymb}
\usepackage{wasysym}

\newcommand{\Ket}[1]{\vert \, #1 \, \rangle}
\newcommand{\Bra}[1]{\langle \, #1 \,\vert}
\newcommand{\MatEl}[3]{\langle \, #1 \,\vert\,#2\,\vert\,#3\,\rangle}

\newcommand{\Avg}[1]{\langle  #1  \rangle}
\newcommand{\la}{\langle}
\newcommand{\ra}{\rangle}

\newcommand{\lp}{\left(}
\newcommand{\rp}{\right)}

\newcommand{\U}{{\cal U}}

\newcommand{\be}{\begin{equation}}
\newcommand{\ee}{\end{equation}}
\newcommand{\bea}{\begin{eqnarray}}
\newcommand{\eea}{\end{eqnarray}}

\renewcommand{\phi}{\varphi}
\renewcommand{\epsilon}{\varepsilon}
\renewcommand{\vec}[1]{{\bf #1}}

\begin{document}

\title{
Phase-Sensitive Probes of %Transverse
Nuclear Polarization in Spin-Blockaded Transport
}
\author{M. S. Rudner$^{1}$, I. Neder$^{1}$, L. S. Levitov$^{2}$, B. I. Halperin$^{1}$}
\affiliation{
$^{(1)}$ Department of Physics,
 Harvard University, 17 Oxford St.,
 Cambridge, MA 02138\\
$^{(2)}$ Department of Physics,
 Massachusetts Institute of Technology, 77 Massachusetts Ave,
 Cambridge, MA 02139}
%% \date{\today}

\begin{abstract}
Spin-blockaded quantum dots provide a unique setting for studying nuclear-spin dynamics in a nanoscale system. Despite recent experimental progress, observing phase-sensitive phenomena in nuclear spin dynamics remains challenging. Here we point out that such a possibility opens up in the regime where hyperfine exchange directly competes with a purely electronic spin-flip mechanism such as the spin-orbital interaction. Interference between the two spin-flip processes, resulting from long-lived coherence of the nuclear-spin bath, modulates the electron-spin-flip rate, making it sensitive to the transverse component of nuclear polarization. In a system repeatedly swept through a singlet-triplet avoided crossing, nuclear precession is manifested in oscillations and sign reversal of the nuclear-spin pumping rate as a function of the waiting time between sweeps. This constitutes a purely electrical method for the detection of coherent nuclear-spin dynamics.

\end{abstract}

\maketitle

%%%%%%% Introduction %%%%%%%%

Due to their long coherence times, nuclear spins offer unique opportunities to study quantum dynamics. %opportunities to study quantum dynamics.
%Nuclear spins, due to their long dephasing times, offer a unique setting to study quantum-coherent dynamics. 
While conventional techniques using RF fields probe macroscopic groups of spins, currently there is wide interest in the local dynamics of nanoscale
groups of spins. 
In particular, semiconducting quantum dots have emerged as a platform for
investigating the intriguing quantum many-body dynamics of
coupled electron and nuclear spins
\cite{HansonRMP2007,Khaetskii2003,Taylor2003,Coish2004,Erlingsson2004,DasSarma2005,Sham2007,Chen2007}.
Many experimental achievements in this field rely on the phenomenon of spin blockade, in which electrons must flip their spins in order to pass through the system in accordance with the Pauli exclusion principle~\cite{OnoScience}. %the Pauli exclusion principle requires 
%In such systems
%In such systems
Such spin-blockaded transport is sensitive to the nuclear-polarization-dependent Overhauser field, which controls the energy splittings between electronic spin states. %couples to the electron spin and shift the electronic energy levels, thus affording 
This sensitivity affords an exciting opportunity to electrically control and detect the states of nuclear spins\cite{Koppens2005,Petta2005}.
% electron transport is sensitive to the nuclear polarization-dependent Overhauser field, which couples to the electron spin and shift the electronic energy levels, thus affording an exciting opportunity to electrically control and detect the states of nuclear spins\cite{Koppens2005,Petta2005}.

% Nevertheless, the goal of probing coherent nuclear phenomena, such as
%Larmor precession or Rabi oscillations, remains unfulfilled. 
%However, %so far such methods have yielded few results probing coherent nuclear phenomena such as Larmor precession or Rabi oscillations.
%in order to probe coherent nuclear phenomena, such as Larmor precession or Rabi oscillations, additional ingredients are needed.
%while this Overhauser %/Zeeman 
%effect probes the magnitudes of the nuclear polarization components parallel and perpendicular to an applied magnetic field, %{\it magnitude} % (absolute value) 
%of nuclear polarization, 
%it provides little information about coherent evolution of the transverse component's {\it orientation}. %, e.g. Larmor precession. % of the {\it orientation} of the 
%transverse component. %their {\it direction}. % of the polarization. 

%It is particularly interesting to study couplings that can be used to electrically probe coherent phenomena, such as Larmor precession or Rabi oscillations, which are not easily probed by the Overhauser effect mentioned above.
In this Rapid Communication, we propose a method which can be used to %by which one can 
electrically probe {\it coherent} nuclear phenomena, such as Larmor precession or Rabi oscillations, which are not easily probed by the Overhauser effect alone.
%% However, although 
%While the abovementioned Overhauser effect can be used to probe the component of nuclear polarization %components 
%of nuclear polarization parallel to an %externally 
%parallel %and perpendicular 
%to an applied magnetic field, 
%it provides little information about coherent precession.  %they provide little information about transverse polarization. 
We identify a regime where transport is sensitive to all three vector-components of the local nuclear polarization, %Bloch vector 
due to coherent interplay between hyperfine coupling and an electron-only spin-coupling such as the spin-orbit interaction\cite{Nowack2007,Pfund2009}. 
As we show, systems with such sensitivity host %can be used to  reveal %a system in this regime may host % with such sensitivity reveals 
a variety of new phenomena arising from coherent nuclear dynamics. 
%The recent observation of %recently, 
%unexpected 
Recently observed oscillations of the nuclear pumping rate controlled by nuclear Larmor precession\cite{Foletti} % was observed \cite{Foletti}, which 
indicate that this unexplored regime in now within experimental reach.

%For concreteness, 
Here we investigate these phenomena in %consider %the system of two electrons in 
a two-electron double quantum
dot in %subjected to 
a uniform magnetic field. % in the Z direction. We focus
%hereafter on the spin-orbit
%interaction\cite{Nowack2007,Pfund2009} as our additional
%electron-only spin coupling, keeping in mind that other
%coupling types, such as e.g. Zeeman coupling to the %spatially
%nonuniform field of a micromagnet\cite{micromagnet}, can
%produce analogous effects. 
The two-electron singlet and $m = \pm 1$ triplet
states are coupled via the
difference of %effective 
hyperfine fields due to transverse
nuclear polarization in the two dots, $\Delta \vec B_{{\rm
nuc},\perp}$ \cite{Koppens2005,Petta2005,JouravlevNazarov}. 
Without an additional coupling, the electron spin-flip rate depends only on  $|\Delta \vec B_{{\rm nuc},\perp}|$, and not on the orientation
of $\Delta \vec B_{{\rm nuc},\perp}$ in the XY plane
\cite{JouravlevNazarov}.

\begin{figure}
\includegraphics[width=3.0in]{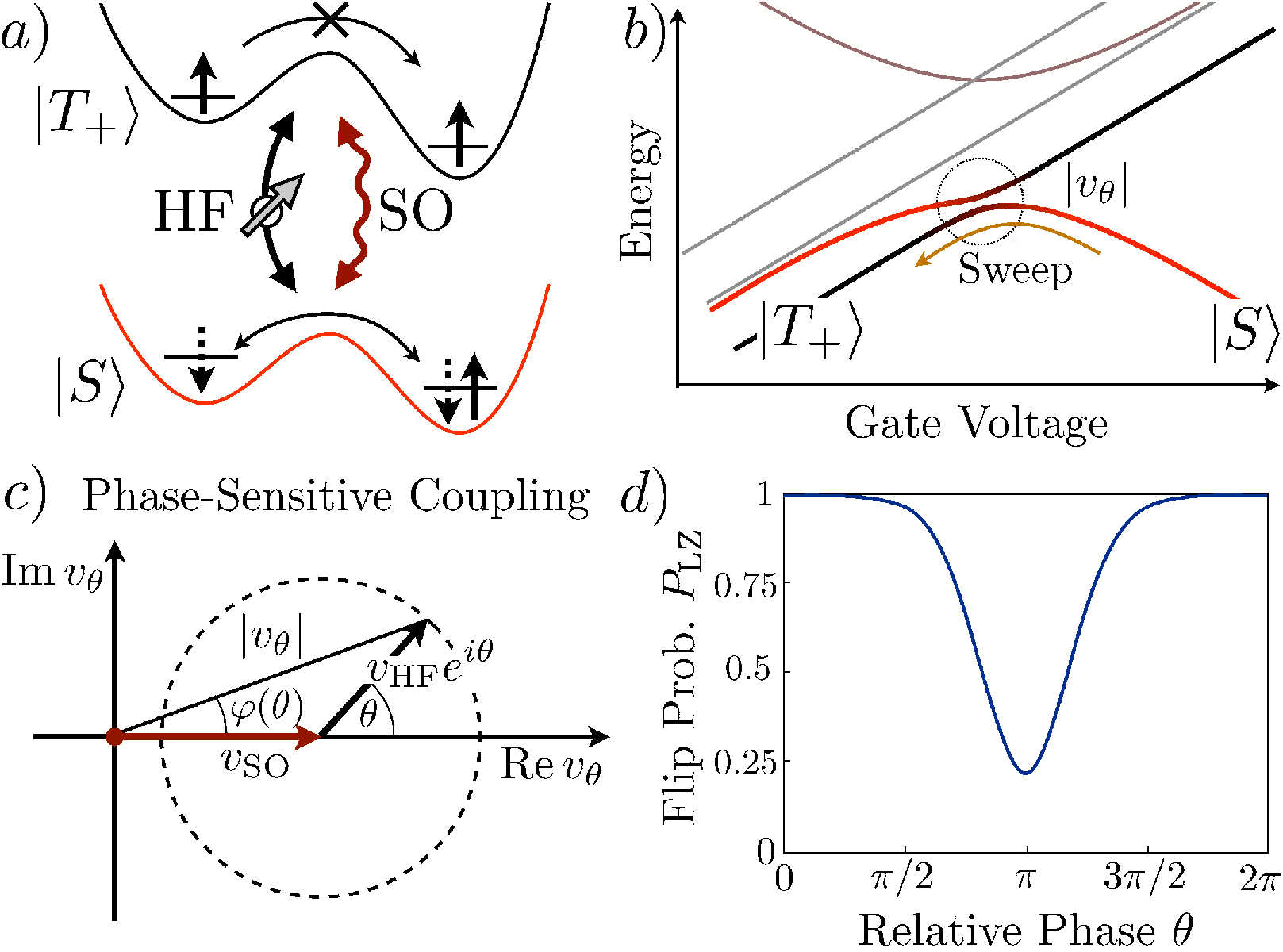}
 \caption[]{(Color online) Coherent interplay of hyperfine and spin-orbit mediated transitions. 
a) Interdot tunneling in the %two-electron 
triplet state is suppressed by Pauli exclusion, but can be mediated by
the hyperfine and spin-orbit interactions which do not conserve electron spin.
b) %Energy levels vs. double dot potential detuning. 
The singlet and triplet levels $\Ket{S}$ and $\Ket{T_+}$ exhibit an avoided crossing with splitting $|v_\theta| = |v_{\rm SO} + v_{\rm HF}e^{i\theta}|$ 
when exchange energy compensates Zeeman energy.
c) Behavior near the $S-T_+$ crossing, controlled by $|v_\theta|$, is sensitive to the relative phase $\theta$ between spin-orbit and hyperfine matrix elements. 
d) Phase-dependent transition probability, Eq.(\ref{P_LZ}), with $v_{\rm SO} = 0.4\sqrt{\hbar\beta}$, and $v_{\rm HF} = 0.6\sqrt{\hbar\beta}$. %, and sweep rate $\beta = 1$ (in units where $\hbar = 1$).
%Destructive interference reduces the transition probability when $\theta \approx \pi$.
}
\label{Setup}
\vspace{-5mm}
\end{figure}
%%%%%%%%%%%%%%%%%%%%%%%%%%%%%%%%%%%%%%%%%%%%%%%%%%%%%%%%%%%%%%%%%%%

The situation becomes considerably more interesting when
singlet-triplet transitions can occur due to either the
spin-orbit or the hyperfine interaction (see Fig.\ref{Setup}).
In this case, we find that the probability of electron
spin-flip depends on \emph{the angle} $\theta$ of the nuclear
polarization measured in the XY plane relative to a fixed axis
determined by the spin-orbit interaction. Such a dependence
makes electron transport sensitive to \emph{the phase} of
nuclear spin precession. 
Below, we analyze this phenomenon for
a Landau-Zener-type process, occurring when the electronic
system is swept through a singlet-triplet level crossing, as
employed e.g. in Refs.\cite{Petta08,Foletti} and depicted in Fig.\ref{Setup}b. 
To make contact with experiment, we focus in particular on
resulting signatures in nuclear spin polarization.
Strikingly, we find robust oscillations in the nuclear spin pumping rate which result from nuclear precession and {\it survive even after averaging over the distribution of random initial nuclear spin states.}
Note that other types of electron-only coupling such as Zeeman coupling to the %spatially
nonuniform field of a micromagnet\cite{micromagnet} can
produce analogous effects. 

% The situation becomes considerably more interesting when singlet-triplet transitions can occur due to either the spin-orbit or the hyperfine interaction (see Fig.\ref{Setup}).
% In this case, we find that the probability of electron spin-flip depends on \emph{the angle} of the nuclear polarization measured in the XY plane relative to a fixed axis determined by the spin-orbit interaction. 
% Such a dependence makes electron transport 
% sensitive to \emph{the phase} of nuclear spin precession, and thus can be used 
% to probe coherent phenomena arising due to nuclear spin dynamics. 

%Below, we analyze this phenomenon for a Landau-Zener-type process, occurring when the electronic system is 
%swept through a singlet-triplet level crossing, as employed e.g. in Refs.\cite{Petta08,Foletti} and depicted in Fig.\ref{Setup}b. To help make contact with experiment, we focus in particular on resulting signatures in nuclear spin pumping.
%Oscillations of the nuclear pumping rate controlled by nuclear Larmor precession, observed in Ref.\cite{Foletti}, indicate that this new regime is now within experimental reach. %%%%%relevant for current and future experiments.

The origin of the spin-angle dependence can be understood heuristically by analyzing the avoided crossing that opens near the degeneracy of
the triplet and singlet levels $\Ket{T_+}$ and $\Ket{S}$ in nonzero magnetic field, circled in Fig.\ref{Setup}b. 
Due to the spin-orbit interaction, tunneling between dots is accompanied by a rotation of electron spin 
\cite{Meir1989} which introduces %gives rise to
a nonzero spin-flip amplitude.
%%%%%%%%%%%%%%%%%
%BEGIN NEW STUFF HERE
%%%%%%%%%%%%%%%
%On the other hand
In addition, the hyperfine interaction between electron and nuclear spins allows %can give rise to 
electronic transitions %between $\Ket{S}$ and $\Ket{T_+}$ %the the electronic states 
accompanied by nuclear spin flips, described by the effective Hamiltonian
%\be
$H_{\rm HF}= A^+ \Ket{S} \Bra{T_+}  +  A^- \Ket{T_+} \Bra{S}$, with $A^{\pm} = \sum_{i=1}^{N_{\rm nuc}} g_i I^{\pm}_i$.
%\ee
%\be
%A^{\pm} = \sum_{i=1}^{N_{\rm Nuc}} g_i I^{\pm}_i  \,
%\ee
%Here $i$ labels individual nuclei, %spins, 
%and the value %magnitude and sign 
The magnitude and sign of each coupling constant $g_i$ depends on the location of nucleus $i$ (positive 
in dot 1 and negative in dot 2). 

Taking advantage of the fact that the number of nuclei interacting with the electrons is very large, $N_{\rm nuc} \approx \mathcal{O}(10^6)$,  %typically of order $10^6$, 
%and that the magnitude of each coupling constant $g_i = \mathcal{O}(1/N_{\rm nuc})$ is small, %of order $1/N_{\rm Nuc}$,
we note that the commutator $[A^+, A^-]$ is typically smaller than $A^+A^-$ by a factor of order $1/N^{1/2}_{\rm nuc}$.
This allows us to treat $A^+$ as a nearly classical variable, 
%Then the commutator $[A^+, A^-]$ will be small, by a factor of order  $1/N_{\rm Nuc}$ compared to $A^{\pm}$ and we may treat $A^+$ as a nearly classical variable, 
$
A^+ \approx v_{\rm HF}\, e^{i\theta}.
$
Here $v_{\rm HF}$ is proportional to the magnitude of the transverse hyperfine difference field 
$\Delta \vec{B}_{{\rm nuc},\perp}$,  
and $\theta$ describes its orientation in the XY plane.

Together, the spin-orbit and hyperfine pieces provide a matrix element between the singlet and triplet states: 
\be
v_\theta\equiv \Bra{S}H\Ket{T_+}=v_{\rm SO}+v_{\rm HF}e^{i\theta} \, .
\label{vtheta} 
\ee
Evolution in the $\{\Ket{S}$, $\Ket{T_+}\}$ subspace 
near the level crossing is described by the $2\times2$ Hamiltonian
\be
H_{ST_+}= \left(
     \begin{array}{cc}
       \frac12 \Delta  & v_\theta^* \\
       v_{\theta}   & -\frac12\Delta
     \end{array} \right)
,\quad \Delta(t)=\epsilon_{T_+}(t)-\epsilon_{S}(t)
,
\label{Ham2x2}
\ee
where $\epsilon_{T_+}$ and $\epsilon_{S}$ are the energies of the diabatic $\Ket{S}$ and $\Ket{T_+}$ states.
The level detuning $\Delta(t)$ can be controlled by  
electrostatic gates and/or magnetic field.

We consider the case where the system is initialized to the ``$(0,2)$" singlet state $\Ket{S}$ with both electrons residing in the right dot (large positive $\Delta$), and then swept through the avoided crossing  to the ``$(1, 1)$'' charge regime with one electron on each dot 
(see Fig.\ref{Setup}b).
For a constant sweep rate $\beta=|d\Delta/dt|$, 
the electron spin flip in such a model 
is interpreted as a Landau-Zener transition occurring with probability (see Fig.\ref{Setup}d)
\be\label{P_LZ}
P_{\rm LZ}=1-\exp\lp-2\pi|v_{\rm SO}+v_{\rm HF}e^{i\theta}|^2/\hbar\beta\rp, 
\ee
where $P_{\rm LZ}$ is the probability to remain in the ground state.
The explicit dependence on the phase $\theta$ of nuclear polarization, which enters through the matrix element $v_\theta$, shows the singlet/triplet transition probability's sensitivity to the transverse nuclear polarization \emph{vector}. 

Model (\ref{Ham2x2}) provides a useful heuristic for understanding electron spin dynamics,
in particular for situations where the nuclear spin state is characterized by a well-defined azimuthal angle $\theta$.
However, to understand the behavior with more general initial states and to account for the effects of back-action on the nuclei, % arising because the hyperfine interaction conserves total spin, 
we must examine the quantum many-body 
dynamics of coherently coupled electron and nuclear spins. 
By solving this problem below, we will further justify the form of Eq.(\ref{vtheta}) and will 
obtain the electron and nuclear spin-flip rates.

\begin{figure}
\includegraphics[width=3.1in]{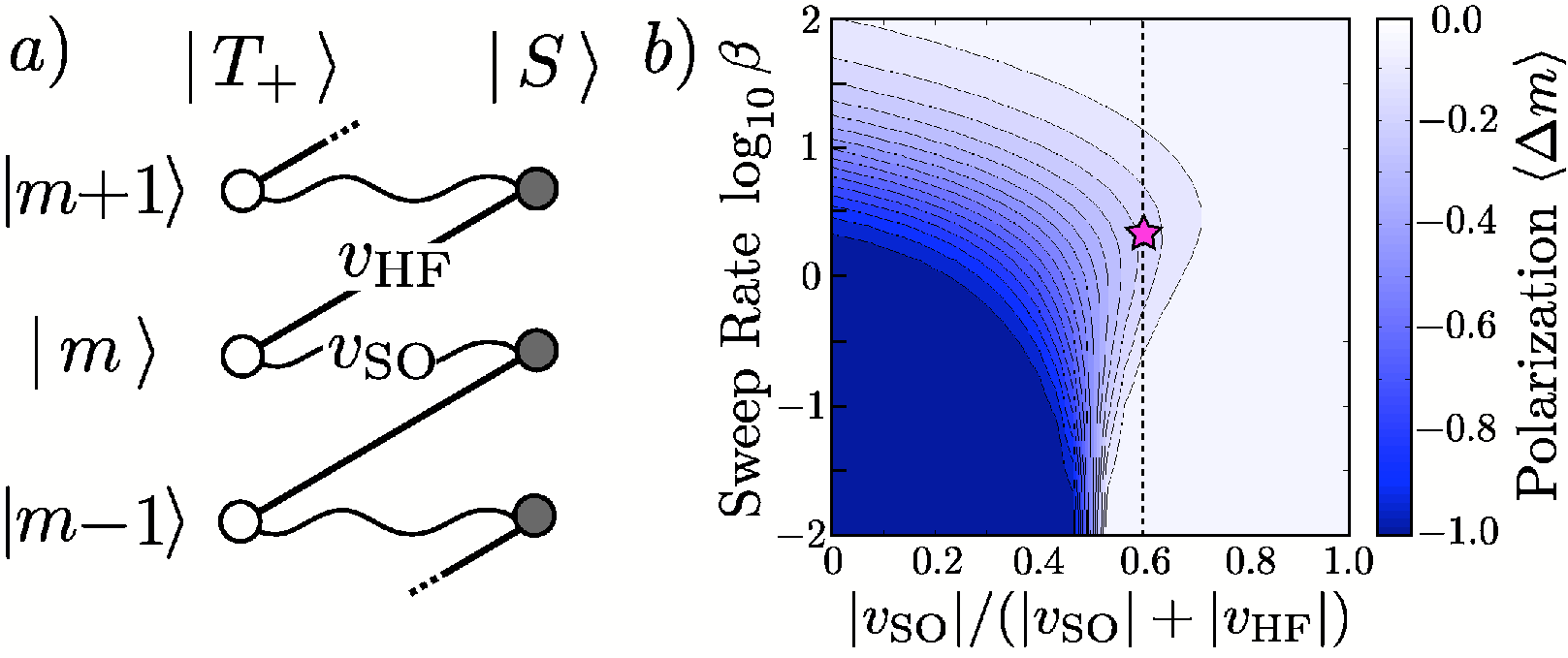}
 \caption[]{(Color online) Quantum walk model of electron-nuclear spin dynamics near the $S-T_+$ crossing, Eq.(\ref{EOM}). 
a) Unit cells are labeled by $m$, the $z$-component of total nuclear spin.  Intracell and intercell hopping correspond to spin-orbit and hyperfine transitions, with matrix elements $v_{\rm SO}$ and $v_{\rm HF}$, respectively.
b) Expected change of nuclear polarization $\Avg{\Delta m}$, Eq.(\ref{DeltaMSS}).
The star indicates the optimal sweep rate $\beta_*$ along the dashed line.
}
\label{QW}
\vspace{-5mm}
\end{figure}

A key new feature here is the relaxed selection rule governing nuclear pumping: %dynamical nuclear polarization: 
due to spin-orbit coupling, electron spin flips may occur \emph{with} or \emph{without} a compensating 
nuclear spin flip. 
Although these two processes apparently lead to orthogonal final states, 
%Because these two processes lead to orthogonal final states, the possibility of interference is surprising. %it is surprising that any interference should be possible.
long-lived coherence of the nuclear spin bath %plays a crucial role by 
allows interference between different transition {\it sequences} (see Fig.\ref{QW}).
% alternative sequences %in which a single electron makes a series of multiple coherent hyperfine and spin-orbit mediated transitions. \mpar{Check here}
Moreover, % as a result, %for appropriate sweep protocols, %in appropriate sequences, we find that 
a \emph{single electron} can change the total nuclear polarization by an  amount $\Delta m$ that can be {\em larger than 1, and of either sign}.% by making a sequence of hyperfine and spin-orbit mediated transitions. 

%By setting $[A^+,A^-] = 0$, we can 
At this point it is convenient to map the problem onto the bipartite 1-dimensional quantum walk shown in Fig.\ref{QW}a, where each unit cell is labeled by $m$, the $z$-projection of the total nuclear spin $I^z  = \sum_i I^z_i$. 
In doing so, $[I^z, A^\pm]$ remains nonzero. %, where $I^z = \sum_i I^z_i$.
%$I^z = \sum_i I^z_i$. %$\vec I= \vec I_1 + \vec I_2$ for dots 1 and 2~
%\footnote{$\vec I_2$ is defined with a 180$^\circ$ rotation such that $\Delta\vec{B}_\perp \propto \vec I_{1,\perp} + \vec I_{2,\perp}$, rather than the conventional $\vec I_{1,\perp} - \vec I_{2,\perp}$.}.
% with  $I^z=I^z_1+I^z_2$, $I^\pm =I^\pm_1-I^\pm_2$.   \mpar{OK?}
%This choice of relative phase  ensures proper symmetry of the hyperfine matrix elements.
Intracell hopping between internal states $T$ and $S$, characterized by $\Delta m = 0$, describes a spin-orbit transition occurring with amplitude $v_{\rm SO}$.
Intercell hopping, characterized by $\Delta m = \pm 1$, describes a hyperfine transition that occurs with the amplitude $v_{\rm HF} = \MatEl{T,\, m-1}{H_{\rm HF}}{S,\, m}$.  
%We choose the basis state phases such that % of the basis states so that  
%$v_{\rm HF},v_{\rm SO} > 0.$  
%The value  of $v_{\rm HF}$ is proportional to the magnitude of $ \Delta \vec B_\perp$. % the difference of the transverse hyperfine fields arising from nuclei on the two dots. 
Initially, we consider sweeps which are fast compared to the nuclear Larmor period, and so neglect the nuclear Zeeman energy.

Because  $N_{\rm{nuc}}$  
is large,  the value of $v_{\rm HF}$ will change very little during the course of a few sweeps, and we treat it here as a constant. % (compare to the discussion in Ref.\cite{WeddingCake}). 
 However, the initial values of  $v_{\rm HF}$ may  differ from one run to another, with a statistical distribution of the form 
$p(v) \propto v e^{-v^2/s^2}$, where $s$ is a constant. 
Additionally, because the typical values of total nuclear spin %on each dot 
will be large, of order $\sqrt {N_{\rm{nuc}}}$, we may take the ladder of allowed values of $m$ to be infinite while the total number of sweeps is not too large.

%%%%%%%%%%%%%%%%%%%%%%%%%%%%%%%%%%%%%%%%%%%%%%%%%%%%%%%%%%%%%%%%%%%
%%%%%%%%%%%%%%%%%%%%%%%%%%%%%%%%%%%%%%%%%%%%%%%%%%%%%%%%%%%%%%%%%%%

The state of the system 
%\mpar{OK?} 
%, described by the amplitudes $\psi^S_m$ and $\psi^T_m$ to find the nuclear spin with $z$-projection $m$ and the electrons in the singlet or triplet state, respectively, 
evolves according to %the equations of motion
\begin{eqnarray}
  \label{EOM} 
\begin{array}{l}
 i\hbar \, \dot{\psi}^T_{m}\ =\ v_{\rm SO}\, \psi^S_{m} \ +\  v_{\rm HF} \psi^S_{m + 1}\ + \ \frac12\Delta(t)\,\psi^T_m\\
 i\hbar \, \dot{\psi}^S_{m}\ =\ v_{\rm SO}\, \psi^T_{m} \  +\ v_{\rm HF} \psi^T_{m - 1}\ - \ \frac12\Delta(t)\, \psi^S_{m}.
\end{array}
\end{eqnarray}

%
%CHANGES END HERE
%
%
%
%%%%%%%%%%%%%%%%%%%%%%%%%%%%%%%%%%%%%%%

At time $t_0$, the system is initialized to the singlet state with polarization $m = m_0$: $\psi^S_m = \delta_{m,m_0}$, $\psi^T_m = 0$. The expected change of polarization, 
\begin{eqnarray}
  \label{DeltaM} \Avg{\Delta m} = \sum_m (m - m_0) P_m;\quad P_m = |\psi_m^S|^2 + |\psi_m^T|^2,
\end{eqnarray}
is determined by the probabilities $\{P_m\}$ to find the nuclear spin with $z$-projection $m$ in the final state at time $t_F$ after the sweep. 
For convenience we take the (non-classical) initial nuclear state to be an eigenstate of $I^z$, but the framework can be applied for more general states. 

%By passing to 
In the Fourier representation $\Ket{\psi_m} = \frac{1}{2\pi}\oint d\theta\, e^{-i m \theta}\,\Ket{\psi_\theta}$, where $\Ket{\psi_m}$ and $\Ket{\psi_\theta}$ are two-component spinors $(\psi_{m}^T,\ \psi_{m}^S)^{\rm T}$ and $(\psi_{\theta}^T,\ \psi_{\theta}^S)^{\rm T}$, the dynamics for different $\theta$ components decouple.  We obtain % we find that the equations of motion are $2\times 2$ block-diagonalized:
\begin{eqnarray}
  \label{EOMk} i \hbar \, \frac{d}{dt}\Ket{\psi_\theta} = H_{ST_+}\Ket{\psi_\theta},
\end{eqnarray}
where $H_{ST_+}$ is given in Eq.(\ref{Ham2x2}).
Here the parameter $\theta$ plays the role of the azimuthal angle of $\Delta\vec{B}_{{\rm nuc},\perp}$ in the classical problem, Eq.(\ref{vtheta}).
Indeed, a state $\psi_\theta^{S,T} \propto \delta(\theta - \theta_0)$ is a coherent state concentrated near $\theta_0$.

To calculate $\Avg{\Delta m}$, we make use of the relation $m\, \Ket{\psi_m} = \frac{i}{2\pi}\oint d\theta \, \frac{d}{d\theta}\left(e^{-i m\theta}\right) \Ket{\psi_\theta}$.
Integrating by parts to move the derivative onto $\Ket{\psi_\theta}$, we find
\begin{eqnarray}
\label{DeltaMk} \Avg{\Delta m} = \frac{1}{2\pi i}\oint d\theta\, \Bra{\psi_\theta}\frac{d}{d\theta}\Ket{\psi_\theta},
\end{eqnarray}
where without loss of generality we have taken $m_0 = 0$.

To evaluate expression (\ref{DeltaMk}), we must solve for the two-level dynamics of $\Ket{\psi_\theta}$ under the time dependent Hamiltonian (\ref{Ham2x2}).
We can write the evolution operator as
\begin{eqnarray}
\label{Uk} U(\theta) =
       \left(
         \begin{array}{cc}
           a_\theta & b_\theta e^{i \phi_0(\theta)}\\
           -b^*_\theta e^{-i\phi_0(\theta)} & a_\theta^*
         \end{array}
       \right),
\ |a_\theta|^2 + |b_\theta|^2 = 1, %\phi_0(\theta) = \arg\,[v_\theta]
\end{eqnarray}
where $\phi_0(\theta) = \arg\,[v_\theta]$ (see Fig.\ref{Setup}c). 
With this parametrization, we have $a_\theta = a_{-\theta}$, and $b_\theta = b_{-\theta}$.
%Here, without loss of generality, we take $v_{\rm SO}$ and $v_{\rm HF}$ to be real.

The initial state $\Ket{\psi_\theta^{(0)}} = (0,\, 1)^{\rm T}$ evolves into $\Ket{\psi_\theta} = (b_\theta e^{i \phi_0(\theta)},\, a^*_\theta)^{\rm T}$.
Because $a_\theta$ and $b_\theta$ are \emph{even} functions of $\theta$, the only term that contributes to $\Avg{\Delta m}$ is that generated when the derivative acts on $e^{i \phi_0(\theta)}$ in Eq.(\ref{DeltaMk}), giving
\begin{eqnarray}
\label{DeltaMSS} \Avg{\Delta m} = -\frac{1}{2\pi}\oint d\theta \lp \frac{d\phi_0}{d\theta}\rp |b_\theta|^2,
\end{eqnarray}
where $|b_\theta|^2$ is the singlet-triplet transition probability in the $\theta$ channel. Comparing this result to that for the net electron spin-flip probability, given by $P_{\rm el}=\frac1{2\pi}\oint |b_\theta|^2 d\theta$, we note that, unlike the situation where only the hyperfine interaction is present, here there is no simple relationship between the electron and nuclear spin-flip rates.

The analysis leading to Eq. (\ref{DeltaMSS}) is valid for \emph{arbitrary} time dependence $\Delta(t)$, and in particular can even describe cases involving multiple $S-T$ crossings.
Remarkably, for very slow sweeps the change in polarization becomes sharply quantized. Indeed, since $|b_\theta|^2\to 1$ in the adiabatic limit, the integral (\ref{DeltaMSS}) is equal to a \emph{winding number}: $\Avg{\Delta m} = -\frac{1}{2\pi}\oint d\phi_0$.
Thus $\Avg{\Delta m} = -1$ or 0 if $v_\theta$ does or does not wind around the origin as $\theta$ traverses the Brillouin zone, depending on the relative magnitudes of $v_{\rm HF}$ and $v_{\rm SO}$\cite{1DPRL}. %  (i.e. if $|v_{\rm HF}| > |v_{\rm SO}|$ or $|v_{\rm HF}| < |v_{\rm SO}|$, {\it cf.} Ref.\cite{1DPRL}).
%\mpar{Ref OK?}
The sign is negative because electron transitions from $S$ to $T_+$ flip nuclear spins from up to down.

The quantity $\Avg{\Delta m}$ exhibits interesting dependence on the sweep speed, which can be analyzed most straightforwardly for linear sweeps $\Delta(t) = -\beta t$, when $|b_\theta|^2$ is given by the Landau-Zener formula (\ref{P_LZ}).
Due to the absence of dynamics for very fast sweeps, and since $\Avg{\Delta m} = 0$ for very slow sweeps in the region $|v_{\rm SO}| > |v_{\rm HF}|$, the polarization efficiency is a \emph{non-monotonic} function of sweep rate (see Fig.\ref{QW}b).
In the limit of weak hyperfine interaction, $|v_{\rm HF}| \ll |v_{\rm SO}|$, expanding the exponential in Eq.(\ref{P_LZ}), we find

\begin{eqnarray}
|\Avg{\Delta m}| \approx (v_{\rm HF}/v_{\rm SO})^2 \lambda e^{-\lambda}
,\quad \lambda =2\pi v^2_{\rm SO}/\hbar\beta .
\end{eqnarray}
%\mpar{rem. exp.}
This expression attains its maximum value $|\Avg{\Delta m}|_{\rm max} = v^2_{\rm HF}/(e v^2_{\rm SO})$ at the optimal sweep rate $\beta_* = 2\pi v^2_{\rm SO}/\hbar$. % ($e = 2.71828...$).
%Interestingly, a similar non-monotonic behavior of the nuclear spin-flip rate was noted in Ref.\cite{Foletti}. 

Next we consider a generalization to multiple sweeps and show that dynamical polarization is sensitive to nuclear Larmor precession between sweeps.
We focus on a sequence of two identical gate sweeps, where each individual sweep is short on the scale of the nuclear Larmor time, but with a waiting time $\Delta t$  between sweeps long enough to allow nuclear spin precession. %that we need to take precession into account.  
Individual sweeps may pass through the $S-T$ crossing one or more times.  
%\mpar{new stuff}
%We consider specifically a sequence of two identical sweeps, and 
Assuming that only one nuclear species contributes to the hyperfine field, %. Then, 
 the change in nuclear polarization due to the two sweeps is given by
\begin{eqnarray} 
\Avg{\Delta m}= \oint \frac{d\theta}{2\pi i}\la \U^{-1}\partial_{\theta}\, \U\ra 
, \ %\nonumber \\
%\quad 
\U = U(\theta+\Delta\theta) \,W  \, U(\theta),
 \label{m(B)}  
\end{eqnarray}
where $U(\theta)$ is the evolution matrix for a single sweep, given by (\ref{Uk}),  $W$ is the electronic evolution operator for the waiting interval $\Delta t$, the expectation value is taken over the state $\Ket{\psi_\theta^{(0)}} = (0,\, 1)^{\rm T}$, and $\Delta\theta=\omega_L\Delta t$ is the precession angle, with $\omega_L$  the nuclear Larmor frequency. %, while. 
Due to the periodicity of $U(\theta)$ in $\theta$, Eq.(\ref{Uk}), $\Avg{\Delta m}$ exhibits periodic oscillations in $\Delta \theta$ which are
a direct manifestation of coherent nuclear polarization dynamics.

The analysis is simple when the detuning $|\Delta|$ is very large in the interval between sweeps. 
Then  $W \approx \exp(i \chi \sigma_3)$, where the phase $\chi$ is large and very sensitive to any fluctuations in the waiting time $\Delta t$ or other parameters of the system.  
The resulting randomness of $\chi$ completely suppresses coherence in the electronic state between sweeps, while nuclei remain coherent.
For demonstration, below we choose the particular protocol $\Delta(t) = \beta (t_n-t)$, for $|t-t_n| < \tau$, and $\Delta =  \infty$ otherwise, where $n= 1,2$, and $\tau \ll \Delta t = t_2 -t_1$ with $t_1$ and $t_2$ the times of passage through the avoided crossing.
%Thus we treat $\chi$ as a random variable, which completely suppresses coherence in the electronic state, but leaves nuclei coherent. % however, and %due to the periodicity of $U(\theta)$ in $\theta$, Eq.(\ref{Uk}), $\Avg{\Delta m}$ will exhibit periodic oscillations in $\Delta \theta$, which are
%a direct manifestation of transverse nuclear polarization dynamics.\mpar{rem. exp.} %; oscillations as a function of $\omega_L \Delta t$ were an important experimental result in  Ref.~\cite{Foletti}.

For this experimentally relevant case, %simple analytic expressions for $\Avg{\Delta m}$ can be derived as follows.
it is helpful to think of the matrix-products defining $\U$ and $\U^{-1}$ in Eq.(\ref{m(B)}) 
as sums over amplitudes associated with all possible histories of $S-T$ transitions. 
Fast electron decoherence between sweeps suppresses the contribution of terms arising from non-identical histories in $\U$ and $\U^{-1}$, leaving behind a sum of classical {\it probabilities} of all trajectories.
Starting with the singlet initial state and summing over all final states, we find 
% $\Avg{\Delta m}=-  (1,1) \,   \partial_\eta \oint \frac{d\theta}{2\pi} M(\theta+\Delta\theta) M(\theta)   \left(\begin{array}{c}\! 0 \! \\ \! 1 \! \end{array}\right)_{\!\eta=0}$
%
\bea
&
\Avg{\Delta m}= \left[-(1,1) \,   \partial_\eta \oint \frac{d\theta}{2\pi} 
 M(\theta+\Delta\theta) M(\theta)  \, (0,1)^{\rm T}\right]_{\eta=0}
%\left(\begin{array}{c}\! 0 \! \\ \! 1 \! \end{array}\right)_{\eta=0}
\nonumber \\
\label{m(B)decohered}
&
M(\theta) \equiv \lp \begin{array}{cc} 1-p(\theta) & p(\theta) e^{\eta \frac{d\phi}{d\theta}}\\ p(\theta) e^{-\eta \frac{d\phi}{d\theta}} & 1-p(\theta) \end{array}\rp
\eea
% 
%where $a_{n,\theta} \equiv \sqrt{1 - p_n(\theta)}$, $b_{n,\theta}e^{i\phi_0(\theta)} \equiv \sqrt{p_n(\theta)}e^{i\phi_n(\theta)}$,  and 
where $\eta$ is an auxiliary variable introduced for bookkeeping, and $p(\theta) = |b_\theta|^2$ is the transition probability in the $\theta$ channel for a single sweep.
%Here we have specialized to the case of symmetric sweeps $\Delta_n(t) = \beta_n t$, $t_0 = -t_F$, %$\Delta(t_0) = -\Delta(t_{F})$, 
%where %equal but opposite contributions to 
%the phase associated with diabatic transitions is zero, ${\rm Im}[a_{n,\theta}] = 0$.
Here the phase $\phi(\theta)$ is the sum of  the geometric part $\phi_0(\theta) = \arg[v_\theta]$, the Stokes phase $\phi_{s}(\theta)$, see e.g. Ref.\cite{Kayanuma}, and the dynamical phase 
$
\phi_{{\rm ad}}(\theta) = \frac{1}{\hbar}\int_{- \tau}^{\tau} E(\beta t, \theta)\, dt
$
associated with adiabatic evolution on a single branch of the two-level spectrum of Hamiltonian $H_{ST_+}$, Eq.(\ref{Ham2x2}), with $E(\Delta, \theta) = -\sqrt{(\Delta/2)^2 + |v_\theta|^2}$.
Equation   (\ref{m(B)decohered}) then gives: %$\Delta\theta = \pi n$, Eq.(\ref{m(B)decohered}) gives:
\bea
\label{EqnDblSwp}
\Avg{\Delta m} = -\oint \frac{d\theta}{2\pi} \left\{ \frac{d\phi}{d\theta}\,p(\theta) + \frac{d\phi'}{d\theta}[ 1 - 2p(\theta)] p'(\theta) \right\}, 
\eea
where $p'(\theta)=p(\theta+\Delta\theta), \phi'(\theta)=\phi(\theta+\Delta\theta)$.

\begin{figure}
\includegraphics[width=2.75in]{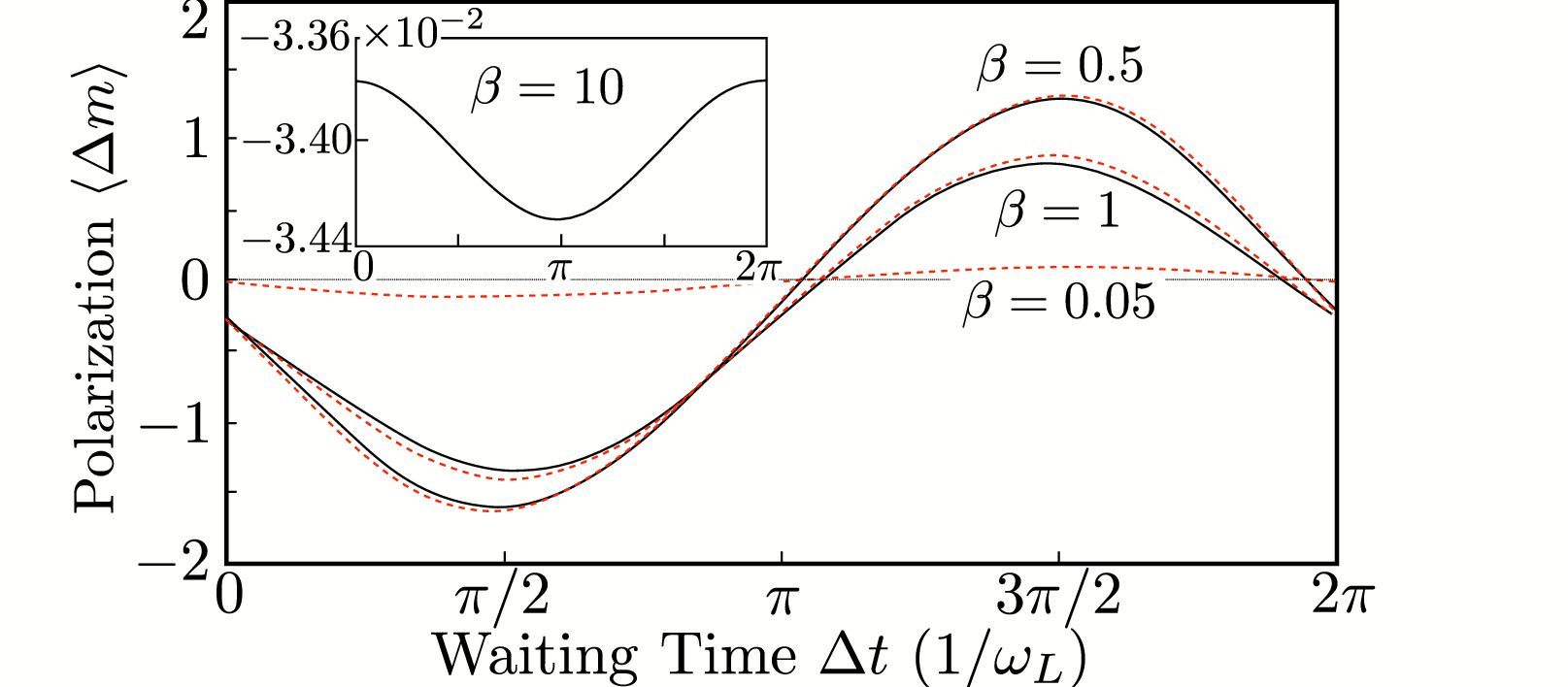}
 \caption[]{(Color online) Polarization due to a round trip through the $S-T_+$ crossing with $ v_{\rm HF} = 0.6, v_{\rm SO} = 0.4, \Delta_0 = 50$, and precession $\Delta\theta = \omega_L\Delta t$ between forward and return sweeps, Eq.(\ref{EqnDblSwp}).
Here $\omega_L$ is the nuclear Larmor frequency and $\Delta t$ is the waiting time.
Dashed lines show $\Avg{\Delta m}$ calculated using $\phi_0(\theta)$ and the dynamical phase, Eq.(\ref{PhiAd}), neglecting Stokes phase.
The oscillation phase shifts by $\pi/2$ for faster sweeps where $\frac{d\phi_{\rm ad}}{d\theta}$ is small (see inset). %is cosine-like for fast sweeps (inset, $\beta = 10$), and becomes sine-like for slower sweeps (main panel).
Units are chosen with $\hbar = 1$.
%The blue dotted line was calculated in this approximation, using $\Delta_0 = 150$.
%Reversed pumping occurs for a range of sweep rates for waiting times equal to an integer number of nuclear Larmor periods (a). No reverse pumping is found for half-integer periods (b).
}
\label{DoubleSweep}
\vspace{-2mm}
\end{figure}

The two terms count the contributions from the two sweeps. %\mpar{Check this}
The factor $1 - 2p(\theta)$ accounts for the fact that a transition from $S$ to $T$  on the  first sweep reverses the direction of nuclear spin flips in the second sweep. % will change the sign of any nuclear spin flip on the second sweep.
Oscillations and sign reversal of $\Avg{\Delta m}$ as a function of the precession angle $\Delta\theta$ are clearly visible in Fig.~\ref{DoubleSweep}, where we have used direct numerical integration to find the transition amplitudes $b(\theta)$ needed to evaluate expression (\ref{EqnDblSwp}). 
%\mpar{rem. comment}

%Note that the direction of nuclear spin pumping can be {\it reversed} for some values of the precession angle between sweeps, providing another hallmark signature of this regime.

%%%%%%%%%%%%%%%%%%%%%%%%%%%%%%%%%%%%%%%%%%%%%%%%%%%%%%%%%%%%%%%%%%%
%%%%%%%%%%%%%%%%%%%%%%%%%%%%%%%%%%%%%%%%%%%%%%%%%%%%%%%%%%%%%%%%%%%

The dynamical phase contributes to $\Avg{\Delta m}$ through  
\be\label{PhiAd}
\frac{d\phi_{\rm ad}}{d\theta} = -\int_{- \tau}^{\tau} \frac{dt}{\hbar} \frac{\partial E(\beta t, \theta)}{\partial \theta}
% \frac{\partial E(\beta t, \theta)}{\partial \theta}
=\frac{4v_{\rm SO}v_{\rm HF}}{\hbar\beta}\sin\theta \sinh^{-1} \frac{|\Delta_0|}{2|v_\theta|},
\ee
%
% \begin{eqnarray}\label{PhiAd}
% \frac{d\phi_{\rm ad}}{d\theta} &=& -\frac{1}{\hbar}\int_{- \tau}^{\tau} dt\, \frac{\partial E(\beta t, \theta)}{\partial \theta}   \nonumber \\
% &=& \frac{ 4 \, v_{{\rm SO}} \,  v_{\rm HF}    \sin\theta}{\hbar\beta}\sinh^{-1} \frac{|\Delta_0|}{2|v_\theta|},
% \end{eqnarray}
%with $\alpha= 4 \, v_{{\rm SO}} \,  v_{\rm HF} /\hbar\beta$.
where $d\phi_{\rm ad}/d\theta$ grows %linearly with the sweep time $2 \tau=2\Delta_0/\beta$ and 
logarithmically with the range of detunings $2\Delta_0$ spanned by the sweep.
Note that $\partial E/\partial \theta$ is the group velocity in the Bloch band corresponding to the 1D quantum walk, Fig.~\ref{QW}a.

This contribution describes DNP buildup at times long after the level crossing.  The in-plane component of electron spin maintains a nonzero average value due to the presence of the spin-orbit interaction.  While this remnant electron spin polarization can be small, its influence on nuclear polarization accumulates over a long time, as indicated by the $1/\beta$ dependence.
As a result, this contribution to DNP typically dominates over that of the geometric phase $\phi_0(\theta)$, see Fig.\ref{DoubleSweep}.

%This dominant contribution for slow sweeps arises from nuclear precession around the transverse component of electron spin, which maintains a nonzero average due to the presence of the spin-orbit interaction (see Fig.~\ref{DoubleSweep}).
%Using the explicit formula for $E(\Delta; \theta)$, we find $d\phi_{\rm ad}/d\theta = \frac{4vv'\sin\theta}{\hbar\beta}\sinh^{-1} \frac{|\Delta(t_0)|}{2|v_\theta|}$, 
%For slow sweeps $|v_\theta|^2/\hbar\beta \gg 1$, the Stokes phase can be neglected compared to the dynamical phase, and 
%
%The net polarization is described well by this simple picture, see Fig.\ref{DoubleSweep}, dashed lines.
%
%Here the oscillation is sine-like, with extrema at odd half-integer multiples of $\pi$; as the sweep rate is increased and the dynamical phase contribution eliminated, the oscillations exhibit a $\pi/2$ phase-shift to become cosine-like. %, albeit with diminished amplitude.

%Similar effects to those described above can occur if there are two species of nuclei present in a system {\it without} spin-orbit coupling.
%Here, however
%There, resonance is determined by the {\it relative} precession phase $\Delta\theta =\Delta\omega\Delta t$, % is an integer multiple of 2$\pi$, 
%where $\Delta\omega$ is the difference of Larmor frequencies.
%\mpar{rev. pumping?}Reversed pumping corresponds to a transfer of polarization from the more weakly to the more strongly coupled species. 

%\mpar{rem.2 species}
%The decoherence procedure used to derive Eq.(\ref{m(B)decohered}) accounts for dephasing during the waiting period between sweeps.
Equation (\ref{m(B)}) can be readily extended to %other sweep protocols,  and to 
arbitrary series of non-identical sweeps. 
%We note that larger values of  $d \phi_{{\rm ad}}/d \theta$, and hence of $|\Avg{\Delta m}|$, can be obtained using a sweep that pauses, for some length of time, at an intermediate value of $|\Delta|$, larger than $  |v_\theta|$, but not too large. 
If we take into account
classical noise in the detuning $\Delta(t)$ during each sweep, the primary effect is to flatten the function $p(\theta)$, causing it to saturate toward the value 1/2 for all $\theta$. %\mpar{check here}
Because the qualitative form of $p(\theta)$ is preserved, however, oscillations resulting from the geometric phase contribution $d\phi_0/d\theta$ survive % our conclusions regarding pumping reversal and oscillations hold 
under quite general circumstances.

Resonant effects at the nuclear Larmor frequency also can be seen directly in the time dependence of electron transport properties.
As the transverse polarization precesses in time, the matrix element $v_\theta$ traces out the dashed circle shown in Fig.\ref{Setup}c, leading to a modulation of $|v_\theta|$ at the nuclear Larmor frequency.
This effect can be detected as an ac modulation of conductance (current) in the system, or as a correlation between the electron spin flip probabilities on successive sweeps through the avoided crossing.
Such measurements would constitute a purely electrical detection of nuclear spin dynamics.

In summary, we have shown that the coherent evolution of nuclear spins in quantum dots can be observed through oscillations and sign reversal of the nuclear spin pumping rate, which occur due to the  coherent interplay between hyperfine and spin-orbit couplings.
%In summary, we have shown that coherent interplay between hyperfine and spin-orbit couplings in %mediated transitions in %spin-blockaded 
%double quantum dots gives rise to new dynamical phenomena which are sensitive to the coherent evolution %{\it transverse} component 
%of nuclear polarization.
%In particular, we find as sign reversal and oscillations of the nuclear spin pumping strength.
Recent observations indicate that this interesting regime is now within experimental reach, opening new possibilities to explore many-body spin dynamics in the solid state.

We thank H. Bluhm, H.-A. Engel, S. Foletti, M. B. Hastings, J. J. Krich,  E. I. Rashba, and A. Yacoby for helpful discussions, and acknowledge 
support from W. M. Keck Foundation Center for Extreme Quantum Information
Theory, and NSF grants DMR-0906475 and DMR-0757145.


\begin{references}
\vspace{-5mm}

\bibitem{HansonRMP2007}
R. Hanson et al., Rev. Mod. Phys. {\bf 79}, 1217 (2007).

\bibitem{Khaetskii2003}
A. V. Khaetskii, D. Loss, and L. Glazman, Phys. Rev. Lett. {\bf 88}, 186802 (2002);
Phys. Rev. B {\bf 67}, 195329 (2003).

\bibitem{Taylor2003}
J. M. Taylor, C. M. Marcus, and M. D. Lukin,
Phys. Rev. Lett. {\bf 90}, 206803 (2003).
%% Long-Lived Memory for Mesoscopic Quantum Bits

\bibitem{Coish2004}
W. A. Coish and D. Loss,
Phys. Rev. B 70, 195340 (2004)
%% Hyperfine interaction in a quantum dot: Non-Markovian electron spin dynamics

\bibitem{Erlingsson2004} S. I. Erlingsson and Y. V. Nazarov, Phys. Rev. B 70, 205327 (2004).

\bibitem{DasSarma2005}
W. M. Witzel, R. de Sousa, S. Das Sarma, Phys. Rev. B {\bf 72}, 161306(R) (2005).
%Quantum theory of spectral-diffusion-induced electron spin decoherence

\bibitem{Sham2007}
W. Yao, R-B. Liu, L. J. Sham, Phys. Rev. B {\bf 74}, 195301 (2006).
%Theory of electron spin decoherence by interacting nuclear spins in a quantum dot

\bibitem{Chen2007}
G. Chen, D. L. Bergman, L. Balents, Phys. Rev. B {\bf 76}, 045312 (2007).


\bibitem{OnoScience}
K. Ono {\it et al.}
%% , D. G. Austing, Y. Tokura, S. Tarucha, 
Science {\bf 297}, 1313 (2002).
%Current Rectification by Pauli Exclusion in a Weakly Coupled Double Quantum Dot System

\bibitem{Petta2005}
%Coherent Manipulation of Coupled Electron Spins in Semiconductor Quantum Dots
J. R. Petta {\it et al.}
%% , A.C. Johnson, J.M. Taylor, E.A. Laird, A. Yacoby, M.D. Lukin, C.M. Marcus, M. P. Hanson, A.C. Gossard, 
Science {\bf 309}, 2180 (2005). 

\bibitem{Koppens2005}
% Control and Detection of Singlet-Triplet Mixing in a Random Nuclear Field
F. H. L. Koppens {\it et al.}
%% , J. A. Folk, J. M. Elzerman, R. Hanson, L. H. Willems van Beveren, I. T. Vink, H. P. Tranitz, W. Wegscheider, L. P. Kouwenhoven, L. M. K. Vandersypen, 
Science {\bf 309}, 1346 (2005). 

\bibitem{Nowack2007}
%Coherent control of a single electron spin with electric fields
K. C. Nowack {\it et al.} %F. H. L. Koppens, Yu. V. Nazarov, L. M. K. Vandersypen, 
Science 318, 1430 (2007).

\bibitem{Pfund2009}
%Dynamics of coupled spins in quantum dots with strong spin-orbit interaction
A. Pfund, I. Shorubalko, K. Ensslin, R. Leturcq, 
Phys. Rev. B {\bf 79}, 121306(R) (2009).

\bibitem{Foletti}
S. Foletti, {\it et al.} %J. Martin, M. Dolev, D. Mahalu, V. Umansky, A. Yacoby, 
arXiv:0801.3613.
%Dynamic nuclear polarization using a single pair of electrons

\bibitem{JouravlevNazarov}
O. N. Jouravlev, Y. V. Nazarov, Phys. Rev. Lett. {\bf 96}, 176804 (2006).

\bibitem{Petta08}
J. R. Petta {\it et al.}
%% , J. M. Taylor, A. C. Johnson, A. Yacoby, M. D. Lukin, C. M. Marcus, M. P. Hanson, A. C. Gossard, 
Phys. Rev. Lett. {\bf 100}, 067601 (2008).
%Dynamic Nuclear Polarization with Single Electron Spins

\bibitem{micromagnet}
M. Pioro-Ladri\`ere {\it et al.}
%, T. Obata, Y. Tokura, Y.-S. Shin, T. Kubo, K. Yoshida, T. Taniyama,  S. Tarucha
Nat. Phys. {\bf 4}, 776 (2008). 


\bibitem{Meir1989}
Y. Meir, Y. Gefen, O. Entin-Wohlman,
Phys. Rev. Lett. {\bf 63}, 798 (1989)



%\bibitem{Johnson2005}
%A. C. Johnson {\it et al.}
%, J. R. Petta, J. M. Taylor, A. Yacoby, M. D. Lukin, C. M. Marcus, M. P. Hanson, A. C. Gossard, 
%Nature {\bf 435}, 925 (2005). 
%Triplet-Singlet Spin Relaxation via Nuclei in a Double Quantum Dot


%\bibitem{Liu08}
%Pauli-spin-blockade transport through a silicon double quantum dot
%H. W. Liu, T. Fujisawa, Y. Ono, H. Inokawa, A. Fujiwara, K. Takashina, Y. Hirayama, Phys. Rev. B {\bf 77}, 073310 (2008).

%\bibitem{Shaji08}
%Spin blockade and lifetime-enhanced transport in a few-electron Si/SiGe double quantum dot
%N. Shaji, C. B. Simmons, M. Thalakulam, L. J. Klein, H. Qin, H. Luo, D. E. Savage, M. G. Lagally, A. J. Rimberg, R. Joynt, M. Friesen, R. H. Blick, S. N. Coppersmith, M. A. Eriksson, Nature Physics {\bf 4}, 540 (2008).

%\bibitem{Pfund}
%Suppression of spin relaxation in an InAs nanowire double quantum dot
%A. Pfund, I. Shorubalko, K. Ensslin, R. Leturcq, Phys. Rev. Lett. {\bf 99}, 036801 (2007). 

%\bibitem{Churchill}
%Transport and Charge Sensing of Spin Blockaded 12C and 13C Nanotube Double Quantum Dots 
%H. O. H. Churchill, A. J. Bestwick, J. W. Harlow, F. Kuemmeuth, D. Marcos, C. H. Stwertka, S. K. Watson, C. M . Marcus, Nature Physics {\bf Need full ref.}

%\bibitem{ISC}
%C. Doubleday Jr., N. J. Turro, J.-F. Wang, Acc. Chem. Res. {\bf 22}, 200 (1989).
%Dynamics of flexible Triplet Biradicals

\bibitem{1DPRL}
%Topological Transition in a Non-Hermitian Quantum Walk
M. S. Rudner and L. S. Levitov, Phys. Rev. Lett. {\bf 102}, 065703 (2009).

\bibitem{Kayanuma}
Y. Kayanuma, Phys. Rev. A {\bf 55}, R2495 (1997).
%Stokes phase and geometrical phase in a driven two-level system

%\bibitem{WeddingCake}
%M. S. Rudner, L. S. Levitov, in preparation.
%Wedding cake long paper

%\bibitem{OnoTarucha}
%Nuclear-Spin-Induced Oscillatory Current in Spin-Blockaded Quantum Dots
%K. Ono and S. Tarucha, Phys. Rev. Lett. {\bf 92}, 256803 (2004). 



%\bibitem{Pfund2009}
%Dynamics of coupled spins in quantum dots with strong spin-orbit interaction
%A. Pfund, I. Shorubalko, K. Ensslin, R. Leturq, Phys. Rev. B {\bf 79}, 121306(R) (2009).


%\bibitem{Engel2001} 
%H.-A. Engel and D. Loss, Phys. Rev. Lett. 86, 4648 (2001)
%% Detection of Single Spin Decoherence in a Quantum Dot via Charge Currents


%\bibitem{Elliot54}
%R. J. Elliot, Phys. Rev. {\bf 96}, 266 (1954).
%Theory of the Effect of Spin-Orbit Coupling on Magnetic Resonance in Some Semiconductors

%\bibitem{Yafet}
%{\bf Need Yafet Reference}

%\bibitem{DyakonovPerel}
%M. I. D'yakonov, V. I. Perel, Sov. Phys. JETP {\bf 33}, 1053 (1971);
%M. I. D'yakonov, V. I. Perel, Sov. Phys. Solid State {\bf 13}, 3023 (1971).
%Spin-orbit induced electron spin relaxation refs

%\bibitem{D-PSpinHall}
%M. I. D'yakonov, V. I. Perel, JETP Lett. {\bf 13}, 467 (1971).
%M. I. D'yakonov, V. I. Perel, Phys. Lett. A {\bf 35}, 459 (1971).
%Original spin-Hall papers 

%\bibitem{Kato2004}
%Y. Kato, R. C. Myers, A. C. Gossard, D. D. Awschalom, Science {\bf 306} 1910 (2004).
%Observation of the Spin Hall Effect in Semiconductors

%\bibitem{Overhauser}
%A. Overhauser, Phys. Rev. {\bf 92}, 411 (1953).
%Polarization of nuclei in metals

%\bibitem{Abragam}
%A. Abragam, Phys. Rev. {\bf 98}, 1729 (1955).
%Overhauser Effect in Nonmetals

%\bibitem{OpticalOrientation}
%{\it Optical Orientation}, edited by F. Meier and B.P. Zakharchenya (Elsevier Science, Amsterdam, 1984).



%\bibitem{RudnerDNP}
%Self-Polarization and Dynamical Cooling of Nuclear Spins in Double Quantum Dots
%M. S. Rudner and L. S. Levitov, Phys. Rev. Lett. {\bf 99}, 036602 (2007).

%\bibitem{RudnerCooling}
%Resonant cooling of nuclear spins in quantum dots
%M. S. Rudner and L. S. Levitov, arXiv:0705.2177.

%\bibitem{Taylor03}
%J. M. Taylor, A. Imamoglu, and M. D. Lukin,
%% Controlling a Mesoscopic Spin Environment by Quantum Bit Manipulation
%Phys. Rev. Lett. {\bf 91}, 246802 (2003).

%\bibitem{1DPRL}
%Topological Transition in a Non-Hermitian Quantum Walk
%M. S. Rudner and L. S. Levitov, Phys. Rev. Lett. {\bf 102}, 065703 (2009).

%\bibitem{Fradkin} Eduardo Fradkin, {\it Field Theories of Condensed Matter Systems},  (Addison-Wesley, 1991).

%\bibitem{HallConductance}
%Quantized Hall Conductance in a Two-Dimensional Periodic Potential
%D. J. Thouless, M. Kohmoto, M. P. Nightingale, M. den Nijs, Phys. Rev. Lett. {\bf 49}, 405 (1982).

%\bibitem{AdiabaticTransport}
%Quantization of particle transport
%D. J. Thouless, Phys. Rev. B {\bf 27}, 6083 (1983); Q. Niu and D. J. Thouless,
%J. Phys. A {\bf 17} 2453 (1984).

%\bibitem{Hatano96} N. Hatano and D. R. Nelson, Phys. Rev. Lett. {\bf 77} 570 (1996);
%% N. Hatano and D. R. Nelson, 
%Phys. Rev. B{\bf  56}, 8651 (1997).
%% Localization transitions in non-Hermitian quantum mechanics

%\bibitem{Efetov97} 
%K. B. Efetov,
%% Directed Quantum Chaos
%Phys. Rev. Lett. {\bf 79}, 491 (1997).

%\bibitem{Beenakker96}
%C. W. J. Beenakker, J. C. J. Paasschens, and P. W. Brouwer,
%% Probability of Reflection by a Random Laser
%Phys. Rev. Lett. 76, 1368 (1996).

%% also: www.nature.com/nphys/journal/v4/n5/full/nphys971.html

%\bibitem{Shnerb98} 
%D. R. Nelson and N. M. Shnerb, Phys. Rev. E{\bf 58}, 1383 (1998).
%% Non-Hermitian Localization and population biology
%% Winding numbers and  Non-Hermitian Localization,
%% Phys. Rev. Lett. 80, 5172 (1998)
%% http://arxiv.org/abs/cond-mat/9801111

%\bibitem{Feinberg97}
%J. Feinberg and A. Zee, Nucl. Phys. B 504, 579 (1997).

%\bibitem{OnoTarucha}
%Nuclear-Spin-Induced Oscillatory Current in Spin-Blockaded Quantum Dots
%K. Ono and S. Tarucha, Phys. Rev. Lett. {\bf 92}, 256803 (2004).

%\bibitem{Koppens}
%Control and Detection of Singlet-Triplet Mixing in a Random Nuclear Field
%F. H. L. Koppens, J. A. Folk, J. M. Elzerman, R. Hanson, L. H. Willems van Beveren, I. T. Vink, H. P. Tranitz, W. Wegscheider, L. P. Kouwenhoven, L. M. K. Vandersypen, Science {\bf 309}, 1346 (2005). 

%\bibitem{Pfund}
%Suppression of spin relaxation in an InAs nanowire double quantum dot
%A. Pfund, I. Shorubalko, K. Ensslin, R. Leturcq, Phys. Rev. Lett. {\bf 99}, 036801 (2007).

%\bibitem{RudnerDNP}
%Self-Polarization and Dynamical Cooling of Nuclear Spins in Double Quantum Dots
%M. S. Rudner and L. S. Levitov, Phys. Rev. Lett. {\bf 99}, 036602 (2007).

%\bibitem{SOvsHF}
%Competition between HF and SO in SB
%M. S. Rudner and L. S. Levitov, in preparation.

%\bibitem{Khaetskii02}
%A. V. Khaetskii, D. Loss, and L. Glazman, 
%Phys. Rev. Lett. {\bf 88}, 186802 (2002).
%% Electron Spin Decoherence in Quantum Dots due to Interaction with Nuclei

%\bibitem{psi^R_k(t)}
%An explicit form of the evolution operator, found from Eq.(\ref{EOMk}), 
%gives $\psi^R_k(t)$ which is nonzero at all $t>0$ 
%except when both $\varepsilon_L = \varepsilon_R$ and $\frac14\gamma < |A_k|$.

%\bibitem{Raikh94}
%T. V. Shahbazyan and M. E. Raikh, 
%Phys. Rev. B {\bf  49}, 17123 (1994).

%\bibitem{Brandes05}
%% Coherent and collective quantum optical effects in mesoscopic systems
%T. Brandes, Phys. Rep. {\bf 408}, 315 (2005). 


%\bibitem{Orellana}
%Dicke effect in a quantum wire with side-coupled quantum dots
%P. A. Orellana, F. Dominguez-Adame, E. Diez, Physica E {\bf 35}, 126 (2006).

%\bibitem{Dicke}
%Coherence in Spontaneous Radiation Processes
%R. H. Dicke, Phys. Rev. {\bf 93}, 99 (1954).

%\bibitem{OnoSB}
%Current Rectification by Pauli Exculsion in a Weakly Coupled Double Quantum Dot System
%K. Ono, D. G. Austing, Y. Tokura, S. Tarucha, Science {\bf 297}, 1313 (2002).

%\bibitem{Koppens}
%Control and Detection of Singlet-Triplet Mixing in a Random Nuclear Field
%F. H. L. Koppens, J. A. Folk, J. M. Elzerman, R. Hanson, L. H. Willems van Beveren, I. T. Vink, H. P. Tranitz, W. Wegscheider, L. P. Kouwenhoven, L. M. K. Vandersypen, Science {\bf 309}, 1346-1350 (2005). 

%\bibitem{Pfund}
%Suppression of spin relaxation in an InAs nanowire double quantum dot
%A. Pfund, I. Shorubalko, K. Ensslin, R. Leturcq, Phys. Rev. Lett. {\bf 99}, 036801 (2007). 

%\bibitem{Churchill}
%Transport and Charge Sensing of Spin Blockaded 12C and 13C Nanotube Double Quantum Dots 
%H. Churchill, APS March Meeting (2008).

%\bibitem{OnoTarucha}
%Nuclear-Spin-Induced Oscillatory Current in Spin-Blockaded Quantum Dots
%K. Ono and S. Tarucha, Phys. Rev. Lett. {\bf 92}, 256803 (2004). 

%\bibitem{RudnerDNP}
%Self-Polarization and Dynamical Cooling of Nuclear Spins in Double Quantum Dots
%M. S. Rudner and L. S. Levitov, Phys. Rev. Lett. {\bf 99}, 036602 (2007).

%\bibitem{RudnerCooling}
%Resonant cooling of nuclear spins in quantum dots
%M. S. Rudner and L. S. Levitov, arXiv:0705.2177.

%\bibitem{Emary}
%Dark states in the magnetotransport through triple quantum dots
%C. Emary, Phys. Rev. B {\bf 76}, 245319 (2007).

%\bibitem{Michaelis}
%All-electronic coherent population trapping in quantum dots 
%B. Michaelis, C. Emary, C. W. J. Beenakker, Eurphys. Lett. {\bf 73} 677 (2006).

%\bibitem{Marquardt}
%Dephasing in sequential tunneling through a double-dot interferometer 
%F. Marquardt and C. Bruder, Phys. Rev. B {\bf 68}, 195305 (2003).



%\bibitem{Fussel}
%Quasimode-projection approach to quantum-dotâ€"photon interactions in photonic-crystal-slab coupled-cavity systems
%D. P. Fussell and M. M. Dignam, Phys. Rev. A {\bf 77}, 053805 (2008).

%\bibitem{Baugh}
% Large Nuclear Overhauser Fields Detected in Vertically Coupled Double Quantum Dots
%J. Baugh, Y. Kitamura, K. Ono, S. Tarucha, Phys. Rev. Lett. {\bf 99}, 096804 (2007).

%\bibitem{CL}
%W. A. Coish and D. Loss, Phys. Rev. B {\bf 72}, 125337 (2005). 
%Singlet-triplet decoherence due to nuclear spins in a double quantum dot

%\bibitem{DelftESR}
%F. H. L. Koppens, C. Buizert, K. J. Tielrooij, I. T. Vink, K. C. Nowack, T. Meunier, L. P. Kouwenhoven, L. M. K. Vandersypen, Nature {\bf 442}, 766 (2006).

%\bibitem{HarvardESR}
%E. A. Laird, C. Barthel, E. I. Rashba, C. M. Marcus, M. P. Hanson, A. C. Gossard, arXiv:0707.0557.

%\bibitem{OpticalOrientation}
%{\it Optical Orientation}, edited by F. Meier and B.P. Zakharchenya (Elsevier Science, Amsterdam, 1984).

%\bibitem{Gammon}
%Electron and Nuclear Spin Interactions in the Optical Spectra of Single GaAs Quantum Dots 
%D. Gammon, {\it et al.}, Phys. Rev. Lett. {\bf 86}, 5176 (2001).

%\bibitem{Abragam}
%A. Abragam, Phys. Rev. {\bf 98}, 1729 (1955).
%Overhauser Effect in Nonmetals

%\bibitem{ENF}
%S. I. Erlingsson, Y. V. Nazarov, V. I. Fal'ko, Phys. Rev. B {\bf 64}, 195306 (2001).
%Nucleus-Mediated Spin-Flip Transitions in GaAs Quantum Dots


\end{references}
\end{document}